\documentclass[nofootinbib,prd,showpacs,twocolumn]{revtex4}
\usepackage{amsmath}
\usepackage{amsfonts}
\usepackage{amssymb}
\usepackage{graphicx}

\begin{document}
\title{Linearized stability analysis of gravastars in
noncommutative geometry}

\author{Francisco S. N. Lobo}
\email{flobo@cosmo.fis.fc.ul.pt} \affiliation{Centro de Astronomia
e Astrof\'{\i}sica da Universidade de Lisboa, Campo Grande, Ed. C8
1749-016 Lisboa, Portugal}

\author{Remo Garattini}
\email{Remo.Garattini@unibg.it}
\affiliation{Universit\`{a} degli Studi di Bergamo,
Facolt\`{a} di Ingegneria, Viale
Marconi 5, 24044 Dalmine (Bergamo) ITALY}
\affiliation{INFN - sezione di Milano, Via Celoria 16, Milan, Italy.}

\date{\today}

\begin{abstract}

In this work, we find exact gravastar solutions in the context of
noncommutative geometry, and explore their physical properties and
characteristics. The energy density of these geometries is a
smeared and particle-like gravitational source, where the mass is
diffused throughout a region of linear dimension $\sqrt{\alpha}$
due to the intrinsic uncertainty encoded in the coordinate
commutator. These solutions are then matched to an exterior
Schwarzschild spacetime. We further explore the dynamical
stability of the transition layer of these gravastars, for the
specific case of $\beta=M^2/\alpha<1.9$, where $M$ is the black
hole mass, to linearized spherically symmetric radial
perturbations about static equilibrium solutions. It is found that
large stability regions exist and, in particular, located
sufficiently close to where the event horizon is expected to form.

\end{abstract}

\maketitle

\section{Introduction}

Recently, an alternative picture for the final state of
gravitational collapse has emerged \cite{Mazur}. The latter,
denoted as a gravastar ({\it grav}itational {\it va}cuum {\it
star}), consists of an interior compact object matched to an
exterior Schwarzschild vacuum spacetime, at or near where the
event horizon is expected to form. Therefore, these alternative
models do not possess a singularity at the origin and have no
event horizon, as its rigid surface is located at a radius
slightly greater than the Schwarzschild radius. More specifically,
the gravastar picture, proposed by Mazur and Mottola \cite{Mazur},
has an effective phase transition at/near where the event horizon
is expected to form, and the interior is replaced by a de Sitter
condensate. This new emerging picture consisting of a compact
object resembling ordinary spacetime, in which the vacuum energy
is much larger than the cosmological vacuum energy, is also
denoted as a ``dark energy star'' \cite{Chapline}. In fact, a wide
variety of gravastar models have been considered in the literature
\cite{gravastar1,gravastar2} and their observational signatures
have also been explored \cite{gravastar3}. In this work, we
consider a further extension of the gravastar picture in the
context of noncommutative geometry. The dynamical stability of the
transition layer of these gravastars to linearized spherically
symmetric radial perturbations about static equilibrium solutions
is also explored. The analysis of thin shells
\cite{linear-thinshell} and the respective linearized stability
analysis of thin shells has been recently extensively considered
in the literature, and we refer the reader to Refs.
\cite{linearstability,linear-WH} for details.

In the context of noncommutative geometry, an interesting
development of string/M-theory has been the necessity for
spacetime quantization, where the spacetime coordinates become
noncommuting operators on a $D$-brane \cite{Witten}. The
noncommutativity of spacetime is encoded in the commutator $\left[
\mathbf{x}^{\mu},\mathbf{x}^{\nu}\right] =i\,\theta^{\mu\nu}$,
where $\theta^{\mu\nu}$ is an antisymmetric matrix which
determines the fundamental discretization of spacetime. It has
also been shown that noncommutativity eliminates point-like
structures in favor of smeared objects in flat spacetime
\cite{Smailagic:2003yb}. Thus, one may consider the possibility
that noncommutativity could cure the divergences that appear in
general relativity. The effect of the smearing is mathematically
implemented with a substitution of the Dirac-delta function by a
Gaussian distribution of minimal length $\sqrt{\alpha}$. In
particular, the energy density of a static and spherically
symmetric, smeared and particle-like gravitational source has been
considered in the following form \cite{Nicolini:2005vd}
\begin{equation}
\rho_{\alpha}(r)=\frac{M}{(4\pi\alpha)^{3/2}}\;\mathrm{exp}\left(
-\frac{r^{2}}{4\alpha}\right)  \,, \label{NCGenergy}
\end{equation}
where the mass $M$ is diffused throughout a region of linear
dimension $\sqrt{\alpha}$ due to the intrinsic uncertainty encoded
in the coordinate commutator.

The Schwarzschild metric is modified when a non-commutative
spacetime is taken into account \cite{Nicolini:2005vd, Esposito}.
The solution obtained is described by the following spacetime
metric
\begin{equation}
ds^{2}=-f(r)\,dt^{2}+\frac{dr^{2}}{f(r)}+r^{2}
\,(d\theta^{2}+\sin^{2}{\theta
}\,d\phi^{2})\,,\label{NCW}
\end{equation}
with $f(r)=1-2m(r)/r$, where the mass function is defined as
\begin{equation}
m(r)=\frac{2M}{\sqrt{\pi}}\gamma\left( \frac{3}{2},\frac
{r^{2}}{4\alpha}\right)  \,,
  \label{massfunction}
\end{equation}
and
\begin{equation}
\gamma\left(  \frac{3}{2},\frac{r^{2}}{4\alpha}\right)
=\int\limits_{0} ^{r^{2}/4\alpha}dt\sqrt{t}\exp\left(-t\right)
\end{equation}
is the lower incomplete gamma function \cite{Nicolini:2005vd}. The
classical Schwarzschild mass is recovered in the limit
$r/\sqrt{\alpha}\rightarrow\infty$. It was shown that the
coordinate noncommutativity cures the usual problems encountered
in the description of the terminal phase of black hole
evaporation. More specifically, it was found that the evaporation
end-point is a zero temperature extremal black hole and there
exist a finite maximum temperature that a black hole can reach
before cooling down to absolute zero. The existence of a regular
de Sitter at the origin's neighborhood was also shown, implying
the absence of a curvature singularity at the origin. Recently,
further research on noncommutative black holes has been
undertaken, with new solutions found providing smeared source
terms for charged and higher dimensional cases \cite{newNCGbh}.
Furthermore, exact solutions of semi-classical wormholes
\cite{Garattini2} in the context of noncommutative geometry were
found \cite{Garattini1}, and their physical properties and
characteristics were analyzed.

This paper is outlined in the following manner. In Section
\ref{sec:II}, we present the generic structure equations of
gravastars, and specify the mass function in the context of
noncommutative geometry. In Section \ref{sec:III}, the linearized
stability analysis procedure is briefly outlined, and the
stability regions of the transition layer of gravastars are
determined. Finally in Section \ref{sec:IV}, we conclude. We adopt
the convention $G=c=1$ throughout this work.

\section{Structure equations of gravastars in noncommutative geometry}
\label{sec:II}

Consider the interior spacetime, without a loss of generality,
given by the following metric, in curvature coordinates
\begin{eqnarray}
ds^2&=&-e^{2\Phi(r)}\,dt^2+\frac{dr^2}{1- 2m(r)/r}
    +r^2 \,d\Omega^2 \label{metric}
\,,
\end{eqnarray}
where $d\Omega^2=(d\theta ^2+\sin ^2{\theta} \, d\phi ^2)$;
$\Phi(r)$ and $m(r)$ are arbitrary functions of the radial
coordinate, $r$. The function $m(r)$ is the quasi-local mass, and
is denoted as the mass function.

The Einstein field equation, $G_{\mu\nu}=8\pi T_{\mu\nu}$ provides
the following relationships
\begin{eqnarray}
m'&=&4\pi r^2 \rho  \label{mass}\,,\\
 \Phi'&=&\frac{m+4\pi
r^3 p_r}{r(r-2m)} \label{Phi}\,,\\
p_r'&=&-\frac{(\rho+p_r)(m+4\pi r^3 p_r)}{r(r-2m)}
+\frac{2}{r}(p_t-p_r)\label{anisotTOV}\,,
\end{eqnarray}
where the prime denotes a derivative with respect to the radial
coordinate. $\rho(r)$ is the energy density, $p_r(r)$ is the
radial pressure, and $p_t(r)$ is the tangential pressure. Equation
(\ref{anisotTOV}) corresponds to the anisotropic pressure
Tolman-Oppenheimer-Volkoff (TOV) equation. The factor $\Phi'(r)$
may be considered the ``gravity profile'' as it is related to the
locally measured acceleration due to gravity, through the
following relationship \cite{gravastar2}: ${\cal
A}=\sqrt{1-2m(r)/r}\,\Phi'(r)$. The convention used is that
$\Phi'(r)$ is positive for an inwardly gravitational attraction,
and negative for an outward gravitational repulsion.

Using the equation of state, $p_r=-\rho$, and taking into account
the field equations (\ref{mass}) and (\ref{Phi}), we have the
following relationship
\begin{equation}
\Phi'(r)=\frac{m- rm'}{r\,\left(r-2m \right)} \,,
            \label{EOScondition}
\end{equation}
which provides the solution given by
\begin{equation}
\Phi(r)=\frac{1}{2}\ln\left[1-\frac{2m(r)}{r}\right] \,.
            \label{EOScondition2}
\end{equation}

One now has at hand three equations, namely, the field Eqs.
(\ref{mass})-(\ref{anisotTOV}), with four unknown functions of
$r$, i.e., $\rho(r)$, $p_r(r)$, $p_t(r)$, and $m(r)$. We shall
consider the approach by choosing a specific choice for a
physically reasonable mass function $m(r)$, thus closing the
system.

In this context, we are interested in the noncommutative geometry
inspired mass function given by Eq. (\ref{massfunction}). The
latter is reorganized into the following form
\begin{equation}
m(r)=\frac{2M}{\sqrt{\pi}}\gamma\left(
\frac{3}{2},\beta\left(\frac {r}{2M}\right)^2\right)  \,,
\label{massfunction3}
\end{equation}
where $\beta$ is defined as $\beta=M^2/\alpha$.

Note that three cases need to be analyzed \cite{Nicolini:2005vd}:
\begin{description}
\item[a)] If $\beta<1.9$, no roots are present;

\item[b)] If $\beta>1.9$, we have two roots,
$r_{-}$ and $r_{+}$, with $r_{+}>r_{-}$;

\item[c)] If $\beta=1.9$, we have $r_{+}=r_{-}$, which may be
interpreted as an {\it extreme} situation, such as the extreme
Reissner-Nordstr\"{o}m metric.
\end{description}

\begin{figure}[h]
\includegraphics[width=2.75in]{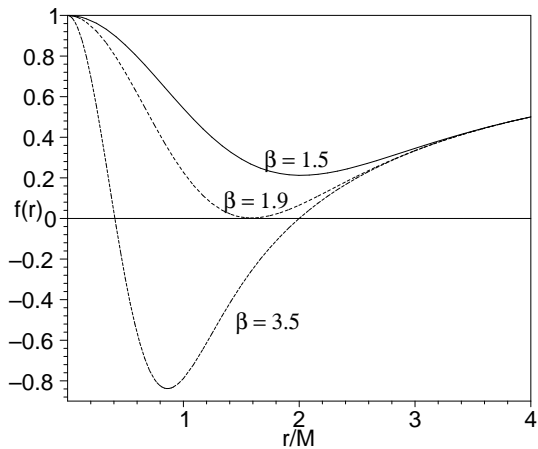}
\caption{The function $f(r)=(1-2m(r)/r)$ is depicted for the three
cases with the following values $\beta=1.5$, $\beta=1.9$ and
$\beta=3.5$, respectively.} \label{fig:mass}
\end{figure}

The function $f(r)=(1-2m(r)/r)$ is depicted in Fig. \ref{fig:mass}
for these three cases for the following values $\beta=1.5$,
$\beta=1.9$ and $\beta=3.5$, respectively. Note that all the roots
lie within the Schwarzschild radius $r_b=2M$, where $M$ is the
total mass of the system.

\section{Linearized stability of gravastars in noncommutative geometry}
\label{sec:III}

\subsection{Junction interface and surface stresses}

We shall model specific gravastar geometries by matching an
interior gravastar geometry, given by Eq. (\ref{metric}), where
the metric functions are given by Eqs. (\ref{EOScondition2}) and
(\ref{massfunction3}), with an exterior Schwarzschild solution
\begin{eqnarray}
ds^2&=&-\left(1- \frac{2M}{r}\right)\,dt^2+\left(1-
\frac{2M}{r}\right)^{-1}\,dr^2
\nonumber \\
&&+r^2 \,(d\theta ^2+\sin ^2{\theta} \, d\phi ^2)
\label{extmetric} \,,
\end{eqnarray}
at a junction interface $\Sigma$, situated outside the event
horizon, $a>r_b=2M$. We emphasize that the larger root $r_+$ lies
inside the Schwarzschild event horizon. Thus, in this work we are
only interested in the case of $\beta<1.9$, which corresponds to
the absence of event horizons for the inner solution.

Consider the junction surface $\Sigma$ as a timelike hypersurface
defined by the parametric equation of the form
$f(x^{\mu}(\xi^i))=0$. $\xi^i=(\tau,\theta,\phi)$ are the
intrinsic coordinates on $\Sigma$, where $\tau$ is the proper time
on the hypersurface. The three basis vectors tangent to $\Sigma$
are given by $e_{(i)}=\partial/\partial \xi^i$, with the following
components $e^{\mu}_{(i)}=\partial x^{\mu}/\partial \xi^i$. The
induced metric on the junction surface is then provided by the
scalar product $g_{ij}=e_{(i)}\cdot e_{(j)}=g_{\mu
\nu}e^{\mu}_{(i)}e^{\nu}_{(j)}$. Thus, the intrinsic metric to
$\Sigma$ is given by
\begin{equation}
ds^2_{\Sigma}=-d\tau^2 + a^2 \,(d\theta ^2+\sin
^2{\theta}\,d\phi^2)  \,.
\end{equation}
Note that the junction surface, $r=a$, is situated outside the
event horizon, i.e., $a>r_b$, to avoid a black hole solution, and
we are only interested in the case of $\beta<1.9$, of Eq.
(\ref{massfunction3}), as emphasized above.

For the specific cases considered in this work, namely, the
interior and exterior spacetimes given by Eqs. (\ref{metric}) and
(\ref{extmetric}), respectively, the four-velocity of the junction
surface $x^\mu(\tau,\theta,\phi)=(t(\tau),a(\tau),\theta,\phi)$ is
given by
\begin{eqnarray}\label{4-velocity}
U_{\pm}^{\mu}=\left(\frac{dt}{d\tau},\frac{da}{d\tau},0,0\right)=
\left(\frac{\sqrt{1-\frac{2m_{\pm}}{a}
+\dot{a}^2}}{1-\frac{2m_{\pm}}{a}},\dot{a},0,0\right),
\end{eqnarray}
where the $(\pm)$ superscripts correspond to the exterior and
interior spacetimes, respectively, so that $m_{\pm}$ are defined
as $m_-=m(a)$ and $m_+=M$, respectively.

The unit normal $4-$vector, $n^{\mu}$, to $\Sigma$ is defined as
\begin{equation}\label{defnormal}
n_{\mu}=\pm \,\left |g^{\alpha \beta}\,\frac{\partial f}{\partial
x ^{\alpha}} \, \frac{\partial f}{\partial x ^{\beta}}\right
|^{-1/2}\;\frac{\partial f}{\partial x^{\mu}}\,,
\end{equation}
with $n_{\mu}\,n^{\mu}=+1$ and $n_{\mu}e^{\mu}_{(i)}=0$. The
Israel formalism requires that the normals point from the interior
spacetime to the exterior spacetime \cite{Israel}. Thus, for the
interior and exterior spacetimes given by the metrics
(\ref{metric}) and (\ref{extmetric}), respectively, the normals
may be determined from Eq. (\ref{defnormal}), or from the
contractions $U^\mu n_\mu=0$ and $n^\mu n_\mu=+1$, and are
provided by
\begin{eqnarray}\label{normals}
n^{\pm}_{\mu}=\left(-\dot{a},\frac{\sqrt{1-\frac{2m_{\pm}}{a}
+\dot{a}^2}}{1-\frac{2m_{\pm}}{a}},0,0\right)\,,
\end{eqnarray}
respectively, with $m_{\pm}$ defined as $m_-=m(a)$ and $m_+=M$, as
before.

The extrinsic curvature is defined as
$K_{ij}=n_{\mu;\nu}e^{\mu}_{(i)}e^{\nu}_{(j)}$. Differentiating
$n_{\mu}e^{\mu}_{(i)}=0$ with respect to $\xi^j$, we have
$n_{\mu}\frac{\partial ^2 x^{\mu}}{\partial \xi^i \,
\partial \xi^j}=-n_{\mu,\nu}\,\frac{\partial x^{\mu}}{\partial
\xi^i}\frac{\partial x^{\nu}}{\partial \xi^j}$, so that the
extrinsic curvature is finally given by
\begin{eqnarray}\label{extrinsiccurv}
K_{ij}^{\pm}=-n_{\mu} \left(\frac{\partial ^2 x^{\mu}}{\partial
\xi ^{i}\,\partial \xi ^{j}}+\Gamma ^{\mu \pm}_{\;\;\alpha
\beta}\;\frac{\partial x^{\alpha}}{\partial \xi ^{i}} \,
\frac{\partial x^{\beta}}{\partial \xi ^{j}} \right) \,.
\end{eqnarray}
Note that, in general, $K_{ij}$ is not continuous across $\Sigma$,
so that for notational convenience, the discontinuity in the
extrinsic curvature is defined as
$\kappa_{ij}=K_{ij}^{+}-K_{ij}^{-}$.

Taking into account the interior spacetime metric (\ref{metric})
and the Schwarzschild solution (\ref{extmetric}), the non-trivial
components of the extrinsic curvature are given by
\begin{eqnarray}
K ^{\tau \;+}_{\;\;\tau}&=&\frac{\frac{M}{a^2}+\ddot{a}}
{\sqrt{1-\frac{2M}{a}+\dot{a}^2}}
\;,  \label{Kplustautau2}\\
K ^{\tau
\;-}_{\;\;\tau}&=&\frac{\frac{m}{a^2}-\frac{m'}{a}+\ddot{a}}
{\sqrt{1-\frac{2m(a)}{a}+\dot{a}^2}} \;, \label{Kminustautau2}
\end{eqnarray}
and
\begin{eqnarray}
K ^{\theta
\;+}_{\;\;\theta}&=&\frac{1}{a}\sqrt{1-\frac{2M}{a}+\dot{a}^2}\;,
 \label{Kplustheta2}\\
K ^{\theta
\;-}_{\;\;\theta}&=&\frac{1}{a}\sqrt{1-\frac{2m(a)}{a}+\dot{a}^2}
\;,  \label{Kminustheta2}
\end{eqnarray}
respectively.

The Einstein equations may be written in the following form,
\begin{equation}
S^{i}_{\;j}=-\frac{1}{8\pi}\,(\kappa ^{i}_{\;j}-\delta
^{i}_{\;j}\kappa ^{k}_{\;k}) \,, \label{Lanczos}
\end{equation}
denoted as the Lanczos equations, where $S^{i}_{\;j}$ is the
surface stress-energy tensor on $\Sigma$. Considerable
simplifications occur due to spherical symmetry, namely $\kappa
^{i}_{\;j}={\rm diag} \left(\kappa ^{\tau}_{\;\tau},\kappa
^{\theta}_{\;\theta},\kappa ^{\theta}_{\;\theta}\right)$. The
surface stress-energy tensor may be written in terms of the
surface energy density, $\sigma$, and the surface pressure, ${\cal
P}$, as $S^{i}_{\;j}={\rm diag}(-\sigma,{\cal P},{\cal P})$. Thus,
the Lanczos equation, Eq. (\ref{Lanczos}), then provide us with
the following expressions for the surface stresses
\begin{eqnarray}
\sigma&=&-\frac{1}{4\pi a} \left(\sqrt{1-\frac{2M}{a}+\dot{a}^2}-
\sqrt{1-\frac{2m}{a}+\dot{a}^2} \, \right)
    \label{surfenergy}   ,\\
{\cal P}&=&\frac{1}{8\pi a} \Bigg[\frac{1-\frac{M}{a}
+\dot{a}^2+a\ddot{a}}{\sqrt{1-\frac{2M}{a}+\dot{a}^2}}
   -\frac{1-\frac{m}{a}-m'+\dot{a}^2 +a\ddot{a}}
{\sqrt{1-\frac{2m}{a}+\dot{a}^2}} \, \Bigg]         \,.
    \label{surfpressure}
\end{eqnarray}

We also use the conservation identity given by
$S^{i}_{j|i}=\left[T_{\mu \nu}e^{\mu}_{(j)}n^{\nu}\right]^+_-$,
where $[X]^+_-$ denotes the discontinuity across the surface
interface, i.e., $[X]^+_-=X^+|_{\Sigma}-X^-|_{\Sigma}$. The
momentum flux term in the right hand side corresponds to the net
discontinuity in the momentum flux $F_\mu=T_{\mu\nu}\,U^\nu$ which
impinges on the shell. The conservation identity is a statement
that all energy and momentum that plunges into the thin shell,
gets caught by the latter and converts into conserved energy and
momentum of the surface stresses of the junction.

Note that $S^{i}_{\tau|i}=-\left[\dot{\sigma}+2\dot{a}(\sigma
+{\cal P} )/a \right]$, so that the conservation identity provides
us with
\begin{equation}
\sigma'=-\frac{2}{a}\,(\sigma+{\cal P})
  \,.\label{consequation2}
\end{equation}
This relationship will be used in the linearized stability
analysis considered below.

\subsection{Linearized stability analysis}

Using the surface mass of the thin shell $m_s=4\pi a^2 \sigma$,
Eq. (\ref{consequation2}) can be rearranged to provide the
following relationship
\begin{equation}
\left(\frac{m_s}{2a}\right)''= \Upsilon -4\pi \sigma'\eta \,,
     \label{cons-equation2}
\end{equation}
with the parameter $\eta$ defined as $\eta={\cal P}'/\sigma'$, and
$\Upsilon $ given by
\begin{equation}
\Upsilon \equiv \frac{4\pi}{a}\,(\sigma+{\cal P}) \,.
\end{equation}

Equation (\ref{cons-equation2}) will play a fundamental role in
determining the stability regions of the respective solutions.
Note that $\eta$ is used as a parametrization of the stable
equilibrium, so that there is no need to specify a surface
equation of state. The parameter $\sqrt{\eta}$ is normally
interpreted as the speed of sound, so that one would expect that
$0<\eta \leq 1$, based on the requirement that the speed of sound
should not exceed the speed of light. We refer the reader to Refs.
\cite{linear-WH} for further discussions on the respective
physical interpretation of $\eta$ lying outside the range $0<\eta
\leq 1$.

Equation (\ref{surfenergy}) may be rearranged to provide the thin
shell's equation of motion given by
\begin{equation}
\dot{a}^2 + V(a)=0 \,.
\end{equation}
The potential is given by
\begin{equation}
V(a)=F(a)-\left[\frac{m_s(a)}{2a}\right]^2
-\left[\frac{aG(a)}{m_s(a)}\right]^2
\,.
\end{equation}
where, for notational convenience, the factors $F(a)$ and $G(a)$
are defined as
\begin{eqnarray}
F(a)=1-\frac{m(a)+M}{a} \quad {\rm and} \quad
G(a)=\frac{M-m(a)}{a} \,.
         \label{factor}
\end{eqnarray}

Linearizing around a stable solution situated at $a_0$, we
consider a Taylor expansion of $V(a)$ around $a_0$ to second
order, given by
\begin{eqnarray}
V(a)&=&V(a_0)+V'(a_0)(a-a_0)
     \nonumber  \\
&&+\frac{1}{2}V''(a_0)(a-a_0)^2+O\left[(a-a_0)^3\right] \,.
\label{linear-potential}
\end{eqnarray}
Evaluated at the static solution, at $a=a_0$, we verify that
$V(a_0)=0$ and $V'(a_0)=0$. From the condition $V'(a_0)=0$, one
extracts the following useful equilibrium relationship
\begin{eqnarray}
\Gamma\equiv\left(\frac{m_s}{2a_0}\right)'
=\left(\frac{a_0}{m_s}\right)\left[
F'-2\left(\frac{a_0G}{m_s}\right)\left(\frac{a_0G}{m_s}\right)'\right]
  \,,
\end{eqnarray}
which will be used in determining the master equation, responsible
for dictating the stable equilibrium configurations.

The solution is stable if and only if $V(a)$ has a local minimum
at $a_0$ and $V''(a_0)>0$ is verified. Thus, from the latter
stability condition, one may deduce the master equation, given by
\begin{equation}
\eta_0 \, \frac{d\sigma^2}{da}\Big|_{a_0} > \Theta\,,
\end{equation}
by using Eq. (\ref{cons-equation2}), where $\eta_0=\eta(a_0)$ and
$\Theta$, for notational simplicity, is defined by
\begin{equation}
\Theta \equiv \frac{1}{2\pi}\left[\sigma \Upsilon+\frac{1}{2\pi
a_0} \left(\Gamma^2-\Psi \right) \right] \,,
       \label{master}
\end{equation}
with
\begin{eqnarray}
\Psi=\frac{F''}{2}-\left[\left(\frac{aG}{m_s}\right)'\right]^2
-\left(\frac{aG}{m_s}\right)\left(\frac{aG}{m_s}\right)'' \,.
\end{eqnarray}

Now, from the master equation we find that the stable equilibrium
regions are dictated by the following inequalities
\begin{eqnarray}
\eta_0 &>& \Omega, \qquad {\rm if} \qquad
\frac{d\sigma^2}{da}\Big|_{a_0}>0\,,      \label{stability1}
       \\
\eta_0 &<& \Omega, \qquad {\rm if} \qquad
\frac{d\sigma^2}{da}\Big|_{a_0}<0\,,       \label{stability2}
\end{eqnarray}
with the definition
\begin{eqnarray}
\Omega\equiv
\Theta\left(\frac{d\sigma^2}{da}\Big|_{a_0}\right)^{-1}\,.
\end{eqnarray}

\subsection{Stability regions}

We now determine the stability regions dictated by the
inequalities (\ref{stability1})-(\ref{stability2}). In the
specific cases that follow, the explicit form of $\Omega$ is
extremely messy, so that we find it more instructive to show the
stability regions graphically.

\begin{figure}[h]
\includegraphics[width=2.75in]{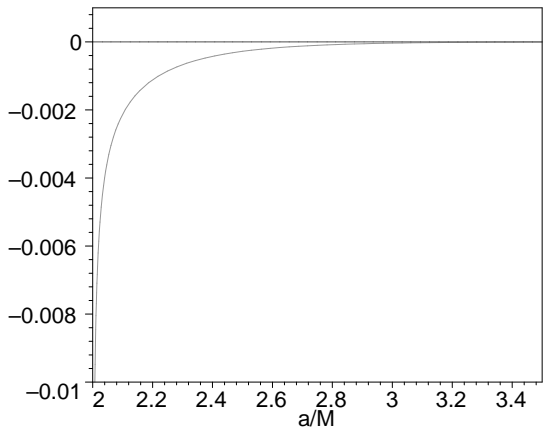}
\caption{Sign of $(d\sigma^2/da)|_{a_0}$, for $\beta=1$, which
corresponds to the absence of horizons.} \label{fig:sigma2}
\end{figure}

For the case of interest under consideration, namely, $\beta<1.9$,
we verify that $d\sigma^2/da|_{a_0}<0$, so that the stability
regions are dictated by inequality (\ref{stability2}). The latter
is shown graphically in Fig. \ref{fig:sigma2}, for the specific
case of $\beta=1.0$.

\begin{figure}[h]
\includegraphics[width=2.75in]{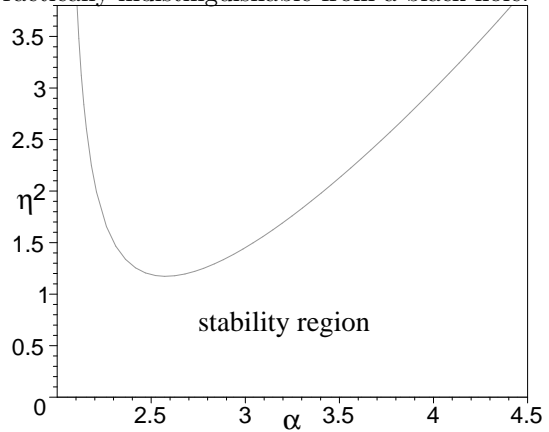}
\caption{Stability plot for $\beta=1$, which corresponds to the
absence of horizons; we have defined $\alpha=a/M$.}
\label{fig:stability}
\end{figure}

Considering the case of $\beta=1.0$, the respective stability
regions are given by the plot depicted below the curve in Fig.
\ref{fig:stability}. Note the existence of large stability regions
sufficiently close to the event horizon. For this case, the
stability regions decrease for increasing $a$, and increases again
as $a$ increases.

The above analysis shows that stable configurations of the surface
layer, located sufficiently near to where the event horizon is
expected to form, do indeed exist. Therefore, considering these
models, one may conclude that the exterior geometry of a dark
energy star would be practically indistinguishable from a black
hole.

\section{Conclusion}
\label{sec:IV}

In this work, we have found exact gravastar solutions in the
context of noncommutative geometry, and briefly explored their
physical properties and characteristics. The energy density of
these geometries is a smeared and particle-like gravitational
source, where the mass is diffused throughout a region of linear
dimension $\sqrt{\alpha}$ due to the intrinsic uncertainty encoded
in the coordinate commutator.

We further explored the dynamical stability of the transition
layer of these dark energy stars to linearized spherically
symmetric radial perturbations about static equilibrium solutions.
It was found that large stability regions do exist, which are
located sufficiently close to where the event horizon is expected
to form, so that it would be difficult to distinguish the exterior
geometry of the gravastars, analyzed in this work, from a black
hole.

\section*{Acknowledgments}

FSNL acknowledges partial financial support of the
Funda\c{c}\~{a}o para a Ci\^{e}ncia e Tecnologia through the
grants PTDC/FIS/102742/2008 and CERN/FP/109381/2009.



\begin{thebibliography}{99}                                                                                               %

\bibitem{Mazur}
P. O. Mazur and E. Mottola, ``Gravitational Condensate Stars: An
Alternative to Black Holes,'' [arXiv:gr-qc/0109035];
%
P. O. Mazur and E. Mottola, ``Dark energy and condensate stars:
Casimir energy in the large,'' [arXiv:gr-qc/0405111];
%
P. O. Mazur and E. Mottola, ``Gravitational Vacuum Condensate
Stars,'' Proc. Nat. Acad. Sci. 111, 9545 (2004)
[arXiv:gr-qc/0407075];
%
G. Chapline, E. Hohlfeld, R. B. Laughlin and D. I. Santiago,
Int. J. Mod. Phys. A 18 3587-3590 (2003).

\bibitem{Chapline}
G. Chapline, ``Dark energy stars,'' [arXiv:astro-ph/0503200];
%
F.~S.~N.~Lobo,
  Class.\ Quant.\ Grav.\  {\bf 23}, 1525 (2006).

\bibitem{gravastar1}
M. Visser and D. L. Wiltshire, Class. Quantum Grav. \textbf{21},
1135 (2004);
%
B. M. N. Carter, Class. Quantum Grav. \textbf{22}, 4551 (2005);
%
P.~Rocha, R.~Chan, M.~F.~A.~da Silva and A.~Wang,
  JCAP {\bf 0811}, 010 (2008);
%
C. Cattoen, T. Faber and M. Visser, Class. Quantum Grav.
\textbf{22}, 4189 (2005);
%
A.~DeBenedictis, D.~Horvat, S.~Ilijic, S.~Kloster and
K.~S.~Viswanathan,
  Class.\ Quant.\ Grav.\  {\bf 23}, 2303 (2006);
%
F.~S.~N.~Lobo and A.~V.~B.~Arellano,
  Class.\ Quant.\ Grav.\  {\bf 24}, 1069 (2007);
%
  N.~Bilic, G.~B.~Tupper and R.~D.~Viollier,
  JCAP {\bf 0602}, 013 (2006);
%
  A.~DeBenedictis, R.~Garattini and F.~S.~N.~Lobo,
  Phys.\ Rev.\  D {\bf 78}, 104003 (2008);
%
  O.~Bertolami and J.~Paramos,
  Phys.\ Rev.\  D {\bf 72}, 123512 (2005);

\bibitem{gravastar2}
F.~S.~N.~Lobo,
  Phys.\ Rev.\  D {\bf 75}, 024023 (2007).

\bibitem{gravastar3} A. E. Broderick and R. Narayan, Class. Quantum Grav.
\textbf{24}, 659 (2007);
%
C. B. M. H. Chirenti and L. Rezzolla, Class. Quantum Grav.
\textbf{24}, 4191 (2007);
%
D. Horvat and S. Ilijic, Class. Quantum Grav. \textbf{24}, 5637
(2007);
%
  V.~Cardoso, P.~Pani, M.~Cadoni and M.~Cavaglia,
  Phys.\ Rev.\  D {\bf 77}, 124044 (2008);
%
C. B. M. H. Chirenti and L. Rezzolla, \prd {bf 78}, 084011 (2008);
%
D. Horvat, S. Ilijic and A. Marunovic, Class. Quantum Grav.
\textbf{26}, 025003 (2009);
%
B. V. Turimov, B. J. Ahmedov and A. A. Abdujabbarov, MPLA, in
press, arXiv:0902.0217 (2009);
%
 T.~Harko, Z.~Kovacs and F.~S.~N.~Lobo,
  Class.\ Quant.\ Grav.\  {\bf 26}, 215006 (2009).


\bibitem{linear-thinshell}
J.~P.~S.~Lemos, F.~S.~N.~Lobo and S.~Quinet de Oliveira, Phys.\
Rev.\ D \textbf{68}, 064004 (2003);
%
F.~S.~N.~Lobo,
Phys.\ Rev.\ D {\bf 75}, 024023 (2007);
%
S.~Sushkov,
Phys. Rev. D {\bf 71}, 043520 (2005);
%
F. S. N. Lobo,
Phys. Rev. D {\bf 71}, 084011 (2005);
%
E.~F.~Eiroa,
Phys. Rev. D {\bf 78}, 024018 (2008);
%
F.~S.~N.~Lobo,
  Class.\ Quant.\ Grav.\  {\bf 21}, 4811 (2004);
%
F.~S.~N.~Lobo,
 Gen.\ Rel.\ Grav.\  {\bf 37}, 2023 (2005);
%
J.~P.~S.~Lemos and F.~S.~N.~Lobo,
  Phys.\ Rev.\  D {\bf 69}, 104007 (2004);
  %
F.~S.~N.~Lobo, arXiv:0710.4474 [gr-qc].

\bibitem{linearstability}
P. R. Brady, J. Louko and E. Poisson,
Phys. Rev. D {\bf 44}, 1891 (1991);
%
M.~Ishak and K.~Lake,
  Phys.\ Rev.\  D {\bf 65}, 044011 (2002);
%
E. F. Eiroa and G. E. Romero
Gen. Rel. Grav. {\bf 36} 651-659 (2004);
%
F. S. N. Lobo and P. Crawford,
Class. Quant. Grav. \textbf{22}, 4869 (2005);
%
F. S. N. Lobo,
Phys. Rev. D {\bf 71}, 124022 (2005);
%
E.~F.~Eiroa and C.~Simeone,
Phys.\ Rev.\  D {\bf 70}, 044008 (2004);
%
E.~F.~Eiroa and C.~Simeone,
Phys.\ Rev.\  D {\bf 71}, 127501 (2005);
%
M.~Thibeault, C.~Simeone and E.~F.~Eiroa,
Gen.\ Rel.\ Grav.\  {\bf 38}, 1593 (2006);
%
F.~Rahaman, M.~Kalam and S.~Chakraborty,
Gen.\ Rel.\ Grav.\ {\bf 38}, 1687 (2006);
%
C.~Bejarano, E.~F.~Eiroa and C.~Simeone,
Phys.\ Rev.\  D {\bf 75}, 027501 (2007);
%
F.~Rahaman, M.~Kalam and S.~Chakraborty,
Int.\ J.\ Mod.\ Phys.\  D {\bf 16}, 1669 (2007);
%
E.~F.~Eiroa and C.~Simeone,
Phys.\ Rev.\  D {\bf 76}, 024021 (2007);
%
F.~Rahaman, M.~Kalam, K.~A.~Rahman and S.~Chakraborty,
Gen.\ Rel.\ Grav.\  {\bf 39}, 945 (2007);
%
M.~G.~Richarte and C.~Simeone,
[arXiv:0711.2297 [gr-qc]].

\bibitem{linear-WH}
E. Poisson and M. Visser,
Phys. Rev. D {\bf 52} 7318 (1995);
%
F.~S.~N.~Lobo and P.~Crawford,
  Class.\ Quant.\ Grav.\  {\bf 21}, 391 (2004).

\bibitem {Witten}E.~Witten,
Nucl.\ Phys.\ B \textbf{460}, 335 (1996);
N.~Seiberg and E.~Witten, JHEP \textbf{9909}, 032 (1999).

\bibitem {Smailagic:2003yb}A.~Smailagic and E.~Spallucci,
J.\ Phys.\ A \textbf{36}, L467 (2003).

\bibitem {Nicolini:2005vd}P.~Nicolini, A.~Smailagic and E.~Spallucci,
Phys.\ Lett.\ B \textbf{632}, 547 (2006).

\bibitem {Esposito}E. Di Grezia, G. Esposito and G. Miele,
J.Phys.A \textbf{41}, 164063 (2008).

\bibitem{newNCGbh}
   R.~Casadio and P.~Nicolini,
   arXiv:0809.2471 [hep-th];
   P.~Nicolini,
   arXiv:0807.1939 [hep-th];
   E.~Spallucci, A.~Smailagic and P.~Nicolini,
arXiv:0801.3519 [hep-th];
   S.~Ansoldi, P.~Nicolini, A.~Smailagic and E.~Spallucci,
   Phys.\ Lett.\  B {\bf 645}, 261 (2007);
   P.~Nicolini,
   J.\ Phys.\ A  {\bf 38}, L631 (2005).

\bibitem {Garattini2}
R.~Garattini and F.~S.~N.~Lobo, Class.\ Quant.\ Grav.\
\textbf{24}, 2401 (2007).

\bibitem {Garattini1}
R.~Garattini and F.~S.~N.~Lobo,
  Phys.\ Lett.\  B {\bf 671}, 146 (2009).

\bibitem{Israel}
K. Lanczos, Ann. Phys. (Leipzig) {\bf 74}, 518 (1924); G. Darmois,
Fasticule XXV ch V (Gauthier-Villars, Paris, France, 1927); W.
Israel, Nuovo Cimento {\bf 44}B, 1 (1966); and corrections in {\it
ibid.} {\bf 48}B, 463 (1966); A. Papapetrou and A. Hamoui, Ann.
Inst. Henri Poincar\'e {\bf 9}, 179 (1968).



\end{thebibliography}
\end{document}